\documentclass[aps,pre,reprint,twocolumn,nofootinbib]{revtex4-1}

\usepackage{subfigure}
\usepackage{amsmath}    % need for subequations
\usepackage{graphicx}   % need for figures
\usepackage{verbatim}   % useful for program listings
\usepackage{color}      % use if color is used in text
\usepackage{subfigure}  % use for side-by-side figures
\usepackage{hyperref}   % use for hypertext links, including those to external documents and URLs
\begin{document}

\title{Signs of universality in the structure of culture}

\author{Alexandru-Ionu\c{t} B\u{a}beanu}
\author{Leandros Talman}
\author{Diego Garlaschelli}
%\affil[1]{Lorentz Institute for Theoretical Physics, Leiden University, The Netherlands}
\affiliation{Lorentz Institute for Theoretical Physics, Leiden University, The Netherlands}
\date{\today}

\begin{abstract}

Understanding the dynamics of opinions, preferences and of culture as whole requires more use of empirical data than has been done so far. 
It is clear that an important role in driving this dynamics is played by social influence, 
which is the essential ingredient of many quantitative models. 
Such models require that all traits are fixed when specifying the ``initial cultural state''. 
Typically, this initial state is randomly generated, from a uniform distribution over the set of possible combinations of traits.
However, recent work has shown that the outcome of social influence dynamics strongly depends on the nature of the initial state.
If the latter is sampled from empirical data instead of being generated in a uniformly random way,
a higher level of cultural diversity is found after long-term dynamics, for the same level of propensity towards collective behavior in the short-term.
Moreover, if the initial state is randomized by shuffling the empirical traits among people, the level of long-term cultural diversity is in-between those obtained for the empirical and uniformly random counterparts.
The current study repeats the analysis for multiple empirical data sets, showing that the results are remarkably similar,
although the matrix of correlations between cultural variables clearly differs across data sets.
This points towards robust structural properties inherent in empirical cultural states, possibly due to universal laws governing the dynamics of culture in the real world. 
The results also suggest that this dynamics might be characterized by criticality and involve mechanisms beyond social influence.

\end{abstract}

\maketitle

\section{Introduction}
\label{Intr}

Quantitative, interdisciplinary research on social systems has recently seen a dramatic increase~\cite{Urry, Lazer},
which is largely motivated by large amounts of data becoming available as a consequence of online and mobile phone activity.
Such data sets allow one to map out large social networks~\cite{Grabowicz, Onnela_1}, consisting of connections and interaction patterns between humans, 
as well as to keep track of how these networks evolve with time~\cite{Ferligo}. 
This stimulated a series of empirical and theoretical studies of the structure and dynamics of social networks~\cite{Wasserman, Easley, Kadushin, Holme}.  
Less attention has been payed to another, complementary aspect of social systems, having to do with the presence and evolution of opinions and preferences:
the structure and dynamics of ``culture''.
This aspect particularly suffers from a lack of empirical research~\cite{Sobkowicz_1},
which is what this article aims at partly compensating for. 
	
This study makes use of quantitative tools developed within an interdisciplinary ``cultural dynamics'' research paradigm, 
which mostly consists of theoretical, model-driven studies, with significant input from physics~\cite{Castellano}.
In addition to embracing the dynamical nature of culture, this paradigm also embraces its multidimensional nature, 
although similar research focusing on single-dimensional dynamics also exists, in which case it is referred to as ``opinion dynamics''~\cite{Castellano} -- 
interesting parallels between opinion dynamics and statistical physics were pointed out already in Ref.~\cite{Galam_1}.
For cultural dynamics, the so-called Axelrod model~\cite{Axelrod} is very representative. 
In this setting, an individual (or agent) is encoded as a sequence of cultural traits (opinions, preferences, beliefs) commonly referred to as a ``cultural vector''.  
Every entry of the vector corresponds to one dimension of culture, also referred to as one ``cultural variable'' or one ``cultural feature''.  
All vectors evolve in time, driven mainly by social influence interactions, along with other ingredients, 
depending on which version of the model is actually used~\cite{Klemm_1, Klemm_2, Kuperman, Flache, Gonzalez-Avella, Centola, Pfau, Battiston, Stivala_2}. 
Any such model requires that all traits of all agents in the initial state are somehow specified,
which is usually done randomly, using a uniform probability distribution over the set of possible cultural vectors -- a uniform ``cultural space distribution''.
This choice is natural if the aim is understanding the (effect of the) dynamics by means of the structure present in the final state, in the absence of any structure in the initial state. 

Taking a somewhat different perspective, Refs.~\cite{Valori, Stivala} explored alternative classes of initial conditions, 
trying instead to understand the effect that the initial state has on the dynamics and on the final state.
It became apparent that the final state is rather sensitive to the initial state. 
In particular, an initial state constructed from an empirical social survey 
behaved significantly different from an initial state that was generated in a uniformly random way~\cite{Valori}.
This implies that cultural dynamics is sensitive to the structure inherent in empirical data.
Such sensitivity is worth exploiting, in order to better understand the empirical structure.
Thus, if the cultural vectors in the initial state correspond to real individuals,
the outcome of social influence models can be used as a quantitative tool for gaining insight about how real individuals are distributed in cultural space,
and indirectly about cultural dynamics in the real world,
since the initial cultural state can be regarded as a partial snapshot of the real world dynamics. 
This is, to a great extent, the perspective of the research presented here, which makes use of a quantitative technique developed in Ref.~\cite{Valori}

On one hand, this technique incorporates the idea of social-influence cultural dynamics, 
which is encoded by a measure of long-term cultural diversity (LTCD), 
which makes use of an Axelrod-type model~\cite{Axelrod} of cultural dynamics with a minimal set of ingredients.
The LTCD quantity estimates the extent to which discrepancies between opinions survive after a long period of cultural dynamics governed by consensus-favoring social influence, 
in the absence of any other process. 
For any given set of cultural vectors (or cultural state), the values of LTCD are shown in correspondence with those of another quantity,
which is a measure of short-term collective behavior (STCB).
The STCB quantity estimates the propensity of the agent population to short-term coordination in terms of their opinions with respect to only one topic.  
This is done using a modification of the Cont-Bouchaud model~\cite{Cont} of social coordination, 
which employs, in a more implicit way, the idea of one-dimensional opinion dynamics driven by social influence, supposedly taking place on a much shorter time-scale.
As described in Sec.~\ref{LTCDaSTCB}, both the LTCD and the STCB quantities are, additionally, functions of the same free parameter, the bounded confidence threshold $\omega$, 
which controls the maximal distance in cultural space for which social influence can operate. 
The common dependence on this parameter is what allows for LTCD to be plotted as a function of STCB.

On the other hand, this technique also incorporates the comparison between the empirical cultural state, 
a uniformly random cultural state and a shuffled one -- 
the latter is constructed by randomly permuting the empirical traits among vectors, thus retaining only part of the empirical information. 
Each of the three cultural states induces, in the LTCD-STCB plot, a curve parametrised by the bounded confidence threshold. 
In Ref.~\cite{Valori}, for the random cultural state, the curve was such that at least one of the two quantities attained a close-to-minimal value for any value of the bounded confidence threshold $\omega$,
meaning that STCB and LTCD were mutually exclusive.
This apparently called for a more complicated description or otherwise suggested a paradox,
since real-world societies seem to allow for both short-term collective behavior and long-term cultural diversity. 
However, for the empirical cultural state,
the two aspects became clearly more compatible, with both quantities attaining intermediate values for a certain $\omega$ interval, 
which appeared a parsimonious way of reconciling LTCD and STCB.
At the same time the shuffled state entailed a compatibility of LTCD and STCB which was intermediate between those obtained for the empirical and random states.

The current study is dedicated to checking the robustness of the LTCD-STCB behavior identified in Ref.~\cite{Valori} across different empirical data sets.
As shown in Sec.~\ref{UnivStrProp}, this behavior appears to be universal, robust across geographical regions and independent of the details of the feature-feature correlation matrix. 
These results are based on multiple sets of cultural vectors, constructed from several empirical sources and examined using the technique briefly described above.
The LTCD and STCB quantities employed by this technique are explained in more detail in Sec.~\ref{LTCDaSTCB}.
Moreover, Sec.~\ref{FormDescCult} gives more details about the formalism behind ``cultural states'' and related concepts. 
Finally, Sec.~\ref{Disc} discusses the results presented throughout the study, possible criticism and questions that can be further investigated.
The manuscript is concluded in Sec.~\ref{Conc}.
Note that, although the definitions in Sec.~\ref{FormDescCult} and Sec.~\ref{LTCDaSTCB} are effectively the same as in Ref.~\cite{Valori},
in view of their importance for this manuscript, they are explained again here from a somewhat different angle, 
while emphasizing certain aspects that previously were only implicit.

\section{The formal representation of culture}
\label{FormDescCult}

The way a cultural state is encoded here is inspired by models of cultural dynamics, in particular by Axelrod-type models~\cite{Axelrod}.
In this paradigm, one deals with a set of variables, called ``cultural features'', which encode information about various properties that individuals can have, 
properties that are inherently subjective and that can change under the action of ``social influence'' arising during person-to-person interactions.
By construction, these variables are allowed to attain only specific values which are here called ``cultural traits''.
The interpretation here is that cultural traits encode ``preferences'', ``opinions'', ``values'' and ``beliefs'' that people can have on various topics,
where each topic is associated to one feature. 

A ``cultural space'' consists of the set of all possible combinations of cultural traits entailed by the set of chosen cultural features, 
together with a measure of dissimilarity between any two combinations. 
Moreover, this dissimilarity, also called the ``cultural distance'', is defined in such a way that it satisfies all the properties of a metric distance (non-negativity, identity of indiscernibles, symmetry and triangle inequality). 
The so-called ``Hamming'' distance is commonly employed for this purpose, which is meaningful as long as there is no obvious ordering of the traits of any feature. 
A cultural space is thus an abstract, discrete, metric space, where each point corresponds to a specific combination of traits.
However, the cultural space is mathematically not a vector space, since there is no notion of additivity attached to it.

A cultural state is essentially the selection of points in the cultural space that needs to be specified for the initial state of cultural dynamics models.
Such a selection is also referred to here as a ``set of cultural vectors'' (SCV), where one ``cultural vector'' is one possible combination of traits.
Formally, this is not a set in the rigorous sense, but a multiset, since it may contain duplicate elements -- identical sequences of traits. 
However, duplicate elements will rarely occur in the initial states constructed for this study, 
since the number of cultural vectors is in practice much smaller than the number of possible points of the cultural space.
On the other hand, they will often occur in the final state. 
This manuscript uses ``SCV'' interchangeably with ``cultural state''.

It is also convenient to consider the notion of ``cultural space distribution'' (CSD), as a discrete probability mass function taking the cultural space as its support.
If the SCV is constructed in a uniformly random way, one implicitly assumes that the underlying cultural space distribution is constant -- all combinations of traits are equally likely. 
If, however, the SCV is constructed from empirical data, the inherent structure may be thought to correspond to non-homogeneities in an underlying CSD, for which the data is representative.

Here, empirical SCVs are mainly constructed from social survey data.
Cultural features are obtained from the questions that are asked in the survey, while the traits of each feature correspond to the possible answers associated to the question.
Thus, a cultural vector represents a sequence of answers that one individual has given to the list of questions in the survey. 
Importantly, a question is selected and encoded as a feature only if it is reasonably subjective,
meaning that it does not ask about demographic or physical aspects concerning the individual (like place of residence, marital status, age), 
and that every allowed answer should be plausible at least from a certain perspective of looking at the question, 
or for people with a certain background or a certain way of thinking.
Moreover, a question is disregarded if the survey is defined in such a way that its list of a-priori allowed answers depends on what answers are given to other questions. 
All features remaining after this filtering -- see Sec.~\ref{AppEmpDat} of the Appendix for more details -- are assumed to contribute equally to the cultural distance, but the way they contribute depends on whether they are treated as nominal or as ordinal variables.
Specifically, the cultural distance $d_{ij}$ between two vectors $i$ and $j$ is computed according to:
\begin{widetext}
\begin{equation}
  \label{CultDist}
  d_{ij} = \frac{1}{F}\sum_{k=1}^{F}\left[ f_{\text{nom}}^k \left(1 - \delta(x_i^k, x_j^k)\right) + (1-f_{\text{nom}}^k) \frac{|x_i^k - x_j^k|}{q^k-1} \right] = \frac{1}{F}\sum_{k=1}^{F} d^k_{ij},
\end{equation}
\end{widetext}
where $F$ is the number of cultural features with $k$ iterating over them, $f_{\text{nom}}^k$ is a binary variable encoding the type of feature $k$ (1 for nominal and 0 for ordinal), 
$q^k$ is the range (number of traits) of feature $k$, $\delta(a,b)$ is a Kroneker delta function of traits $a$ and $b$ (of the same feature) and $x_i^k$ is the trait of cultural vector $i$ with respect to feature $k$.
This definition reduces to the Hamming distance in case there are only nominal variables present.
The second equality sign gives a formulation of the cultural distance as a sum over feature-level cultural distance contributions $d^k_{ij}/F$.

These feature-level contributions allow one to formulate, following Ref.~\cite{Valori}, a notion of feature-feature covariance: 
\begin{equation}
	\sigma^{k,l} = \frac{\langle d_{ij}^k d_{ij}^l \rangle_{i,j \in \overline{1,N}}^{i < j} - \langle d_{ij}^k \rangle_{i,j \in \overline{1,N}}^{i < j} \langle d_{ij}^l \rangle_{i,j \in \overline{1,N}}^{i < j}}{F^2}
\end{equation}
valid for any two features $k$ and $l$, regarldess of $f_{\text{nom}}^k$ and $f_{\text{nom}}^l$. 
Note that the averaging is performed over all $N(N-1)/2$ distinct pairs $(i,j), i \neq j$ of cultural vectors, rather than over all $N$ cultural vectors. 
The feature-feature covariances can be used to define the associated feature-feature (Pearson) correlations via:
\begin{equation}
	\label{FFCor}
	\rho^{k,l} = \frac{\sigma^{k,l}}{\sqrt{\sigma^{k,k}\sigma^{l,l}}}
\end{equation}
which measures the extent to which large/small distances in terms of feature $k$ are associated to large/small distances in terms of feature $l$.
One can definitely see the $F\times F$ correlation matrix $\rho$ as a reflection of a CSD that is compatible with the data.
In general, however, the correlation matrix will only retain part of the information encoded in the CSD, 
first because $\rho^{k,l}$ retains only part of the information in the 2-dimensional contingency table of features $k$ and $l$,
second because a CSD is essentially an F-dimensional contingency table, which might entail all kinds of higher-order correlations.

Assuming the definition of cultural distance given by Eq.~\eqref{CultDist}, a cultural space is already specified by the list of features taken from an empirical data set, together with the associated ranges and types.
In this empirically-defined cultural space, it is meaningful to talk about several types of SCVs.
First, an empirical SCV is constructed from the empirical sequences of traits of the individuals selected from those sampled by the survey. 
Second, a shuffled SCV is constructed by randomly permuting the empirical traits among individuals, independently for every feature. 
Third, a random SCV is constructed by randomly choosing the trait of every person, for every feature.
Note that the shuffled SCV exactly reproduces, for each feature, the empirical frequency of each trait, 
while disregarding all information about the frequencies of co-occurrence of various combinations of traits of two or more different features.
Thus, shuffling destroys all feature-feature correlations $\rho^{k,l}$, as well as any higher-order correlations entailed by the empirical SCV, retaining only the information encoded in the marginal probability distributions associated to individual features.
On the other hand, a random SCV retains nothing of the information inherent in the empirical SCV. 

Finally, note that the mathematical definition of cultural distance illustrated by Eq.~\eqref{CultDist}, already used in Refs.~\cite{Valori} in~\cite{Stivala}, is neither unique nor very sophisticated. 
Other definitions might capture differences in opinions, preferences, values, beliefs, attitudes and associated behavior tendencies in better, more precise ways -- see Ref.~\cite{Castner} for a sophisticated approach. 
However, the current definition is arguably good enough for the problems explored in this study and for how they are attacked. 

\section{Long-term cultural diversity and short-term collective behavior}
\label{LTCDaSTCB}

This section focuses on two quantities that are evaluated on sets of cultural vectors, namely the LTCD and STCB quantities mentioned above.
These are based on the ideas of cultural and opinion dynamics, respectively, driven by social influence in a population of interacting agents
-- as explained below, multidimensional cultural dynamics is explicitly implemented in LTCD, while unidimensional opinion dynamics is implicitly implemented in STCB.
Each agent is associated to one of the cultural vectors in the SCV that is studied. 
For simplicity, both quantities assume that there is no physical space nor a social network that would constrain the interactions between agents. 
In both cases, the interactions are assumed to only be constrained by how the agents are distributed in cultural space. 
Specifically, only if the distance between two cultural vectors is smaller than the bounded confidence threshold $\omega$ are the two agents able to influence each other's opinions in favor of local consensus: there needs to be enough similarity between the cultural traits of two people if any of them is to convince the other of anything. 
This picture is inspired by assimilation-contrast theory~\cite{Sherif}, Ref.~\cite{Flache} being the first study that explicitly uses the bounded-confidence threshold in the context of cultural dynamics, 
after having already been in use in the context of opinion dynamics for some time -- see Ref.~\cite{Lorenz} for an overview.
The bounded confidence threshold $\omega$ functions like a free parameter on which both the LTCD and the STCB quantities depend, for any given SCV.

The LTCD quantity is a measure of the extent to which the given SCV favors cultural diversity on the long term, 
namely a survival of differences in cultural traits at the macro level, in spite of repeated, consensus-favoring interactions at the micro level. 
In the real world, boundaries between populations belonging to different cultures appear to be resilient with respect to social interactions across them~\cite{GG, Barth, Boyd}.
The measure relies on a Axelrod-type model~\cite{Axelrod} of cultural evolution with bounded confidence, which is applied on the SCV. 
This is meant to computationally simulate the evolution of cultural traits under the action of dyadic social influence, in the absence of other processes that may be present in reality. 
According to this model, at each moment in time, two agents $i$ and $j$ are randomly chosen for an interaction. 
If the distance $d_{ij}$ between their cultural vectors is smaller than the threshold $\omega$, then, with a probability proportional to $1-d_{ij}$, 
for one of the features that distinguishes between the two vectors, one of the agents changes its trait to match the other.
With time, agents become more similar to those that are within a distance $\omega$ in the cultural space.
The dynamics stops when several groups are formed, within which agents are completely identical to each other, but too dissimilar across groups for any trait-changing interaction to occur. 
These groups are called ``cultural domains'', term formulated in the context of the original Axelrod model~\cite{Axelrod}, which also included a physical/geographical, 2-dimensional lattice but no (explicit) bounded confidence threshold.
The normalized number of such cultural domains for a given value of $\omega$, averaged over multiple runs of the model, defines the LTCD quantity:
\begin{equation}
 \text{LTCD}(\omega) = \frac{\langle N_D \rangle_{\omega}}{N},
\end{equation}
where $N_D$ is the cultural domains in the final (or absorbing) state of this model, the normalization being made with respect to $N$, the size of the SCV.

The STCB quantity is a measure of the extent to which the given SCV favors collective behavior (or social coordination) on the short term, 
namely the extent to which the agents associated to the cultural vectors in the set would, due to social influence, tend to take actions or make choices in a similar, coordinated way rather than independently from each other.
Bursts of fashion and popularity\cite{Onnela_2, Ratkiewicz, Fortunato}, rapid diffusion of rumors, gossips and habits\cite{Castellano, Chakrabarti} and speculative bubbles and herding behavior on the stock markets\cite{Cont, Sinha} are real-world examples of collective behavior on the short term.
The measure relies on a Cont-Bouchaud type model~\cite{Cont}, which deals with an aggregate choice or opinion of the entire agent population on one issue, which for simplicity is assumed here to be represented by a binary variable, which could encode, for instance, liking vs disliking an item.
According to the model, when collectively confronted with this issue, the agents within a connected group effectively make the same choice or express the same opinion.
In this context (where physical space and social network are disregarded), a connected group is a subset of agents that form a connected component in the graph obtained by introducing a link for every pair $(i,j)$ of agents that are culturally close enough to socially influence each other $d_{ij} < \omega$.
Based on this approximation, the aggregate, normalized choice of the entire population is expressed as a weighted average over the choices of the connected components,
where the weight of the $A$th component is the size $S_A$ of this component. 
However, the group choices themselves are still assumed to be binary, equiprobable random variables with values $\{-1,+1\}$.
Thus, the aggregate, normalized choice is also a random variable, but one that is non-uniformly distributed over some set of rational numbers within $[-1,1]$, 
in a manner that depends on the set of group sizes $\{S_A\}_{\omega}$ induced by a specific value of the $\omega$ threshold.
The spread of this aggregate probability distribution provides the coordination measure that defines the STCB. 
It turns out that this quantity can be analytically computed, for a given $\omega$, according to~\cite{Valori}:
\begin{equation}
  \text{STCB}(\omega) = \sqrt{\sum_A \left( \frac{S_A}{N} \right)_{\omega}^2},
\end{equation}
where the summation is carried over the cultural connected components labeled by different $A$ values.
Note that only the sizes $S_A$ of the components enter the calculation, 
which are in turn determined by the cultural graph obtained by thresholding the $d_{ij}$ matrix by $\omega$.
Also note that STCB is higher when the agents are more concentrated in fewer and larger components. 

There is a crucial difference between the LTCD and the STCB measures:
while the former assumes that agents move in cultural space under the action of social influence, 
the latter assumes that the agents remain fixed in cultural space while they make their decision on one issue which is external to the cultural space. 
Although the STCB implicitly assumes that social influence occurs within the cultural components, 
this influence is supposedly too superficial and too short-lived too also alter the cultural vectors themselves. 
Thus, the LTCD and STCB quantities are concerned with two different time-scales: 
a long time-scale for which cultural vectors and distances are dynamic and a short time-scale for which cultural vectors and distances are fixed. 
Moreover, while LTCD requires computer simulations, the STCB is computed in an analytical way.
Thus, LTCD can be seen as a characteristic of the final cultural state resulting from a long, cultural dynamics process, 
while the STCB can be can be seen as a property of the initial cultural state.

\begin{figure*}
\centering
	\subfigure[][]{\includegraphics[width=5.5cm]{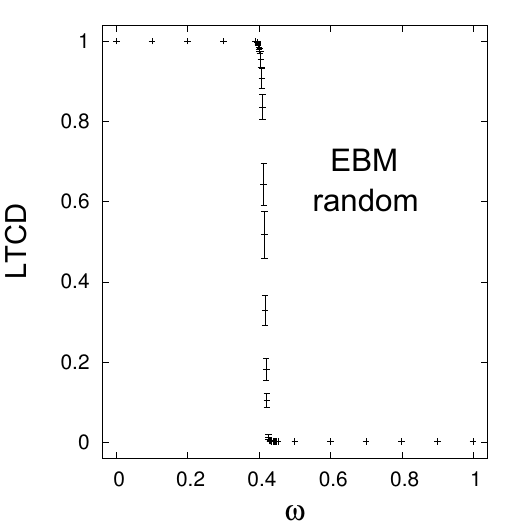}\label{FigRand-D}}
	\subfigure[][]{\includegraphics[width=5.5cm]{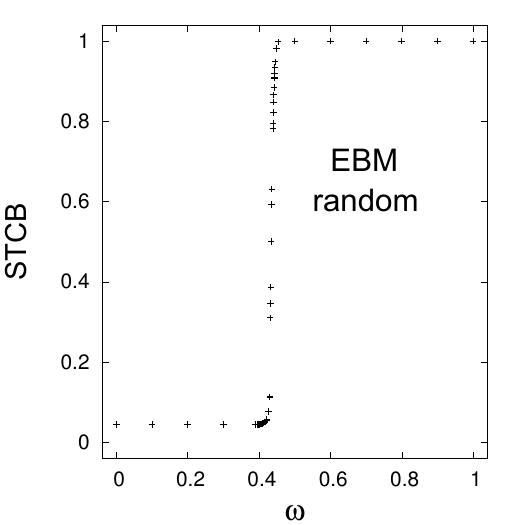}\label{FigRand-C}} 
	\subfigure[][]{\includegraphics[width=5.5cm]{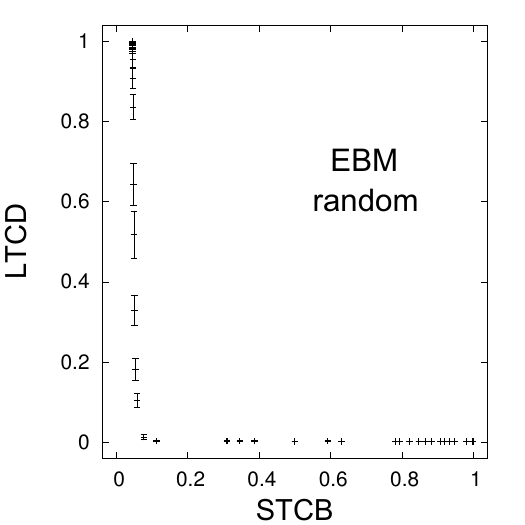}\label{FigRand-DvC}}
	\caption{The interplay between long-term cultural diversity and of short-term collective behavior for a random set of cultural vectors. 
	Showing the LTDC$(\omega)$ dependence \protect\subref{FigRand-D}, the STCB$(\omega)$ dependence \protect\subref{FigRand-C} and the $\omega$-induced LTDC-STCB correspondence \protect\subref{FigRand-DvC},
	for a random set of $N=500$ cultural vectors, in the cultural space of the Eurobarometer (EBM) data set (see Sec.~\ref{UnivStrProp}). 
}
\label{FigRand}
\end{figure*}

It is worth explicitly illustrating, with Fig.~\ref{FigRand}, the behavior of the LTCD and the STCB quantities for a random SCV.
The SCV is defined with respect to the cultural space of one of the data sets introduced in Sec.~\ref{UnivStrProp}. 
Figs.~\ref{FigRand-D} and~\ref{FigRand-C} show, respectively, the dependence of the LTCD and STCB measures on the bounded-confidence threshold $\omega$,
while Fig.~\ref{FigRand-DvC} shows the correspondence between the LTCD and STCB measures obtained by eliminating $\omega$. 
The same data points are used for all 3 plots, where each point records all the 3 quantities (LTCD, STCB and $\omega$).  
The LTCD quantity is averaged, for each point, over 10 runs of the cultural dynamics model, with the associated standard deviations shown by the error bars.

Fig.~\ref{FigRand-D} shows that LTCD decreases with $\omega$: for large $N$, LTCD goes from $1$ to $0$ as $\omega$ goes from $0$ to $1$.
This is doe to $\omega$ controlling the range of interaction in the cultural space.
In general, convergence of agents happens in parallel in several regions of the cultural space, towards several points that are out of range of each other. 
Thus, $\omega$ also controls the expected number of such convergence points, which in turn determines the expected number of cultural domains in the final state and thus the LTCD value --
the latter three quantities decrease with increasing $\omega$. 
If $\omega$ is small enough, there is effectively no successful interaction and thus no movement in cultural space, so each agent ``converges'' to one, distinct point (assuming that all vectors are different from each other in the initial state). 
If $\omega$ is large enough, all agents tend to converge to the same  point in the cultural space. 
Note that, in terms of $\omega$, these two extreme cases are actually two regimes, separated by a sharp decrease of LTCD over some intermediate $\omega$ interval.
This sharp decrease can actually be understood as an order-disorder phase transition, where the disordered phase corresponds to low $\omega$, while the ordered phase corresponds to high $\omega$.
This type of transition has been previously studied in the context of the Axelrod model~\cite{Castellano_2, Battiston}, although in terms of a differently defined control parameter -- 
the (average) feature range $q$ rather than the bounded-confidence threshold $\omega$. 

Fig.~\ref{FigRand-C} shows that STCB is decreasing with $\omega$: in the limit of large $N$, STCB goes from $0$ to $1$ as $\omega$ goes from $0$ to $1$.
This is due to $\omega$ controlling the extent to which agents are culturally connected to each other. 
Higher $\omega$ implies fewer, but larger connected components in the cultural graph, thus a higher predisposition for coordination.  
If $\omega$ is small enough, there is one connected component for every agent, while if $\omega$ is small enough, there is one connected component containing all agents.
Similarly to above, these two cases correspond to two regimes separated by a sharp increase of STCB, 
which can be again understood as a phase transition -- it is actually a symmetry breaking phase transition, as explained in Ref.~\cite{Valori}. 
  
Fig.~\ref{FigRand-DvC} shows that, as $\omega$ increases, one goes from the upper-left corner (high LTCD, low STCB) to the lower-right corner (low LTCD, high STCB), 
by first passing through the lower-left corner (low LTCD, low STCB). 
In other words, the sharp decrease of LTCD happens before the sharp increase of STCB, meaning that the critical $\omega$ of the LTCD phase transition is lower than that of the STCB phase transition.
This is also visible at a close, comparative inspection of Figs.~\ref{FigRand-D} and~\ref{FigRand-C}.
The $\omega$-region for which both the LTCD and the STCB attain low values corresponds to a special situation for which 
there is a relatively high level of convergence in the final cultural state (low LTCD), in spite of a relatively low level of connectivity in the initial cultural state (low STCB).
This is apparently explained by the fact that movement in cultural space at a certain point in the cultural dynamics simulation facilitates further movement that would not have been possible at an earlier moment,
so it is enough to have a few pairs of agents that can initially influence each other to gradually set a large fraction of the other agents in motion and in the end achieve a large amount of convergence. 
In any case, Fig.~\ref{FigRand-DvC} shows that at least one of the two quantities has to attain a close-to-minimal value, regardless of the bounded-confidence threshold $\omega$.

According to the considerations above, long-term cultural diversity and short-term collective behavior seem to be mutually exclusive, suggesting a paradox~\cite{Valori}, at least if one accepts that real socio-cultural systems allow for both aspects. 
However, the above calculations make use of a random SCV, which assumes that the underlying cultural space distribution is uniform. 
Ref.~\cite{Valori} showed that an empirical SCV allows for much more compatibility, with both quantities attaining intermediate values for a certain $\omega$ interval --
as shown in Sec.~\ref{UnivStrProp}, this translates to a higher LTCD-STCB curve than the one shown in Fig.~\ref{FigRand-DvC} -- meaning that the apparent paradox is solved by using realistic data about cultural traits.
Moreover, a shuffled SCV entails a compatibility level that is in between those entailed by a random and by an empirical SCV.
Thus, Ref~\cite{Valori} showed that an empirical SCV has enough structure to dramatically affect the behavior of social-influence dynamics acting upon it, aspect which had been neglected in the past.

\section{Results}
\label{UnivStrProp}

\begin{figure*}
\centering
	\subfigure{\includegraphics[width=8cm]{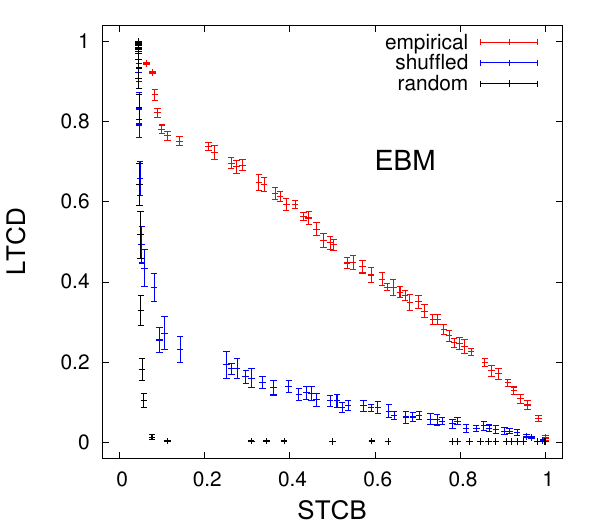}}
	\subfigure{\includegraphics[width=8cm]{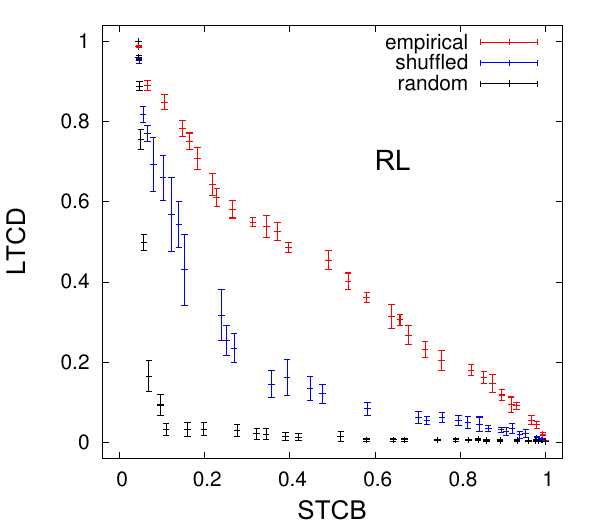}} \\ \vspace{-0.3cm}
	\subfigure{\includegraphics[width=8cm]{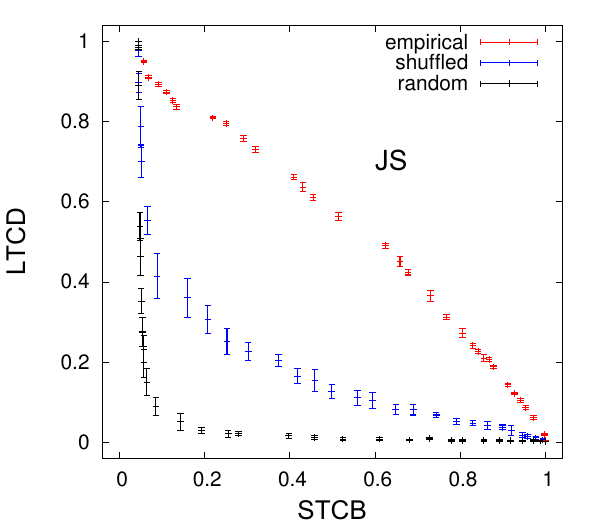}}
	\subfigure{\includegraphics[width=8cm]{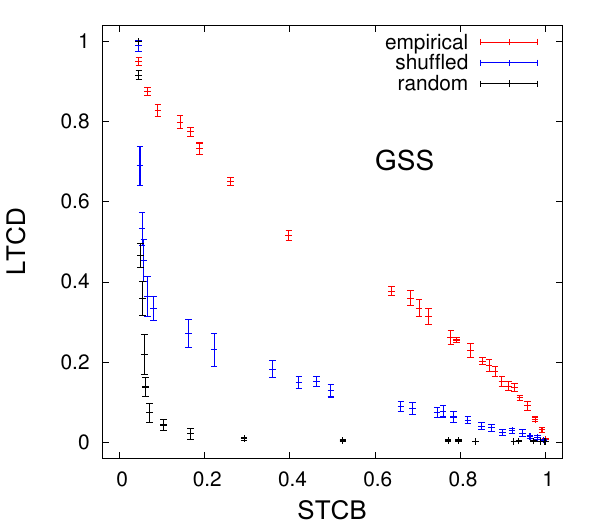}}
	\caption{The correspondence between long-term cultural diversity (LTCD) and short-term collective behavior (STCB) for the empirical (red), shuffled (blue) and random (black) sets of cultural vectors,
  for four data sets: Eurobarometer (EBM), General Social Survey (GSS), Religious Landscape (RL) and Jester (JS). 
  Error bars denote standard deviations over multiple cultural dynamics runs.
  There are $N=500$ elements in each set of cultural vectors. 
}
\label{FigEmpirDvC}
\end{figure*} 

The findings of Ref.~\cite{Valori} are based on one data set. 
It is important to understand whether the observed properties are in fact robust across different populations and across different topics. 
This is accomplished by repeating the analysis of Ref.~\cite{Valori} on four data sets.
These are taken from different sources, thus containing different cultural features and recording the traits of different people. 
The four data sources are:
the Eurobarometer (EBM), containing opinions on science, technology and various European policy issues of people in EU countries~\cite{EBM};
the General Social Survey (GSS), containing opinions on a great variety of topics of people in the US~\cite{GSS};
the Religious Landscape (RL), containing religious beliefs and attitudes on certain political issues of people in the US~\cite{RL};
Jester, containing online ratings of jokes~\cite{JS}. 

\begin{figure*}
\centering
	\subfigure[][]{\includegraphics[width=5.5cm]{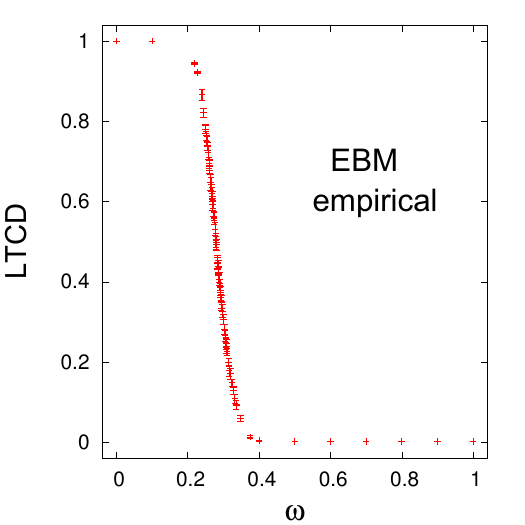}\label{FigEmpir-D}}
	\subfigure[][]{\includegraphics[width=5.5cm]{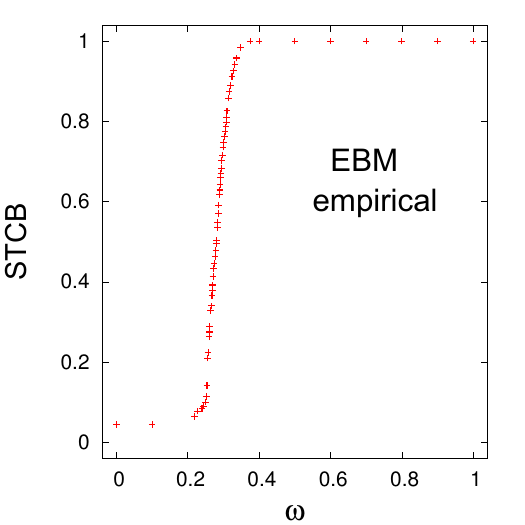}\label{FigEmpir-C}} 
	\subfigure[][]{\includegraphics[width=5.5cm]{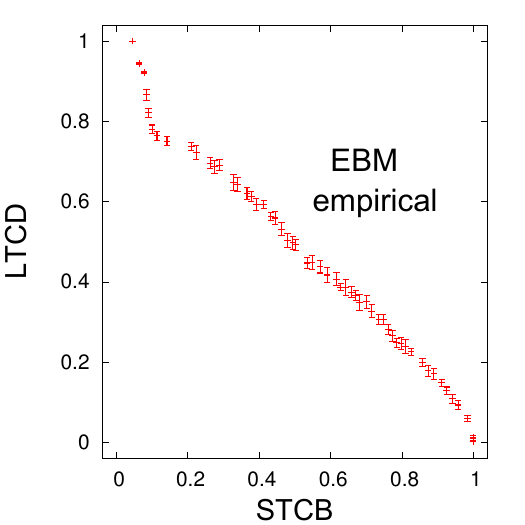}\label{FigEmpir-DvC}}
	\caption{The interplay between long-term cultural diversity and short-term collective behavior for an empirical set of cultural vectors. 
	Showing the LTDC$(\omega)$ dependence \protect\subref{FigEmpir-D}, the STCB$(\omega)$ dependence \protect\subref{FigEmpir-C} and the $\omega$-induced LTDC-STCB correspondence \protect\subref{FigEmpir-DvC},
	for an empirical set of $N=500$ cultural vectors, constructed from the Eurobarometer (EBM) data set. 
}
\label{FigEmpir}
\end{figure*} 

Fig.~\ref{FigEmpirDvC} suggests that the properties highlighted by the LTCD-STCB curves are indeed universal. 
The 4 panels correspond to the 4 empirical data sets that are used. 
In each panel, the 3 curves correspond to the 3 levels of preserving the empirical information: 
full information (red), corresponding to the empirical SCV; 
partial information (blue), corresponding to the shuffled SCV; 
no information (black), corresponding to the random SCV.
Note that, for every data set, the empirical SCV allows for more compatibility between LTCD and STCB than the shuffled SCV, which in turn allows for more compatibility than the random SCV.
Also note that the empirical LTCD-STCB correspondence is always close to the second diagonal.
These qualitative observations constitute the basis for the claim of there being universal structural properties underlying empirical sets of cultural vectors.

\begin{figure*}
\centering
	\subfigure{\includegraphics[width=8cm]{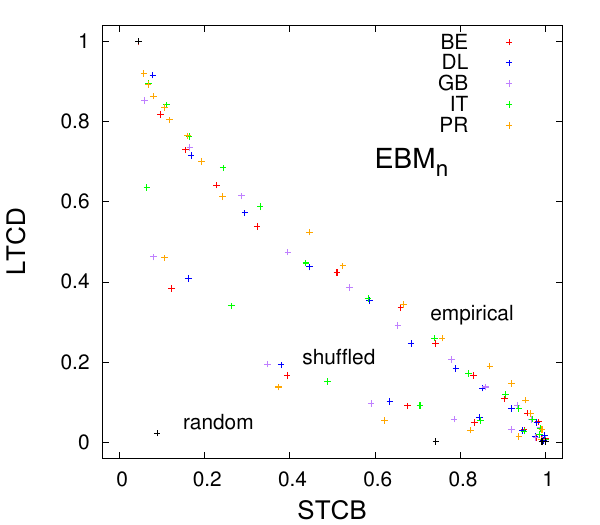}}
	\subfigure{\includegraphics[width=8cm]{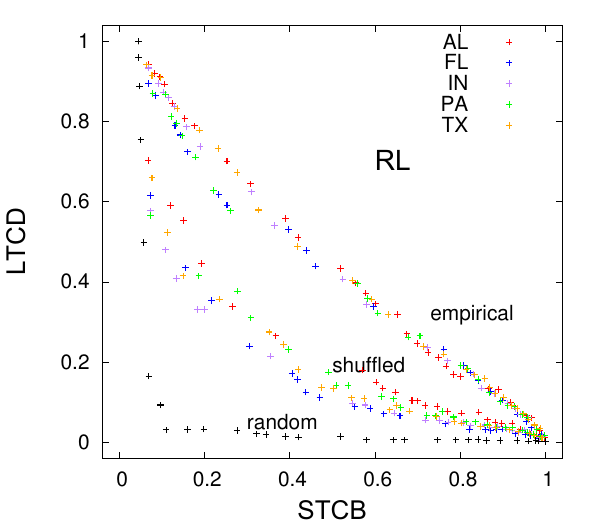}}
	\caption{
  The correspondence between long-term cultural diversity (LTCD) and short-term collective behavior (STCB) for empirical and shuffled sets of cultural vectors
  constructed from country-level and state-level samples of Eurobarometer-nominal ($\text{EBM}_{\text{n}}$) data (left) and Religious Landscape (RL) data (right) respectively.
  There are $N=500$ elements in each set of cultural vectors.
  For visual clarity, error bars are omitted and the same colors are used for both the empirical and shuffled cases,
  while the LTCD-STCB curve is also shown for one random set of cultural vectors in each case.
}
\label{FigEmpirDvC_Geo}
\end{figure*} 

In relation to aspects discussed at the end of Sec.~\ref{FormDescCult}, the change of the LTCD-STCB curve when going from the random to the shuffled and further to the empirical CSV visible in Fig.~\ref{FigEmpirDvC} is related to the LTCD phase transition coming closer to the STCB phase transition.
As $\omega$ increases, for the random case, the LTCD phase transition is almost over when the STCB phase transition begins, 
for the shuffled case there is more overlap between the high-$\omega$ part of the former and the low-$\omega$ part of the latter,
while for the empirical case there is an almost perfect overlap between the two.
The empirical behavior is illustrated by Fig.~\ref{FigEmpir}:
within the $\omega\in[0.2,0.4]$ interval, the decrease in LTCD is systematically accompanied by an increase in STCB.
If one accepts that real-world systems are favorable for both LTCD and STCB and that the respective quantities used here are defined in a sensible way, 
this reasoning suggests that real-world systems function close to criticality, from the perspective of both measures:
only at criticality or close to it are both quantities allowed to attain non-vanishing values in the empirical case.   
In order to stay away from criticality, the system would need to abandon either the propensity towards LTCD or the propensity to STCB.
This suggests, as a speculation or conjecture, that the concept of self-organized criticality~\cite{Bak} might actually play an important role in a complete theory of cultural dynamics. 
If this is correct, then a complete theory of cultural dynamics should have no need of fine-tuning the $\omega$ parameter.

\begin{figure*}
\centering
	\hspace{0.5cm}
	\subfigure{\includegraphics[width=8cm]{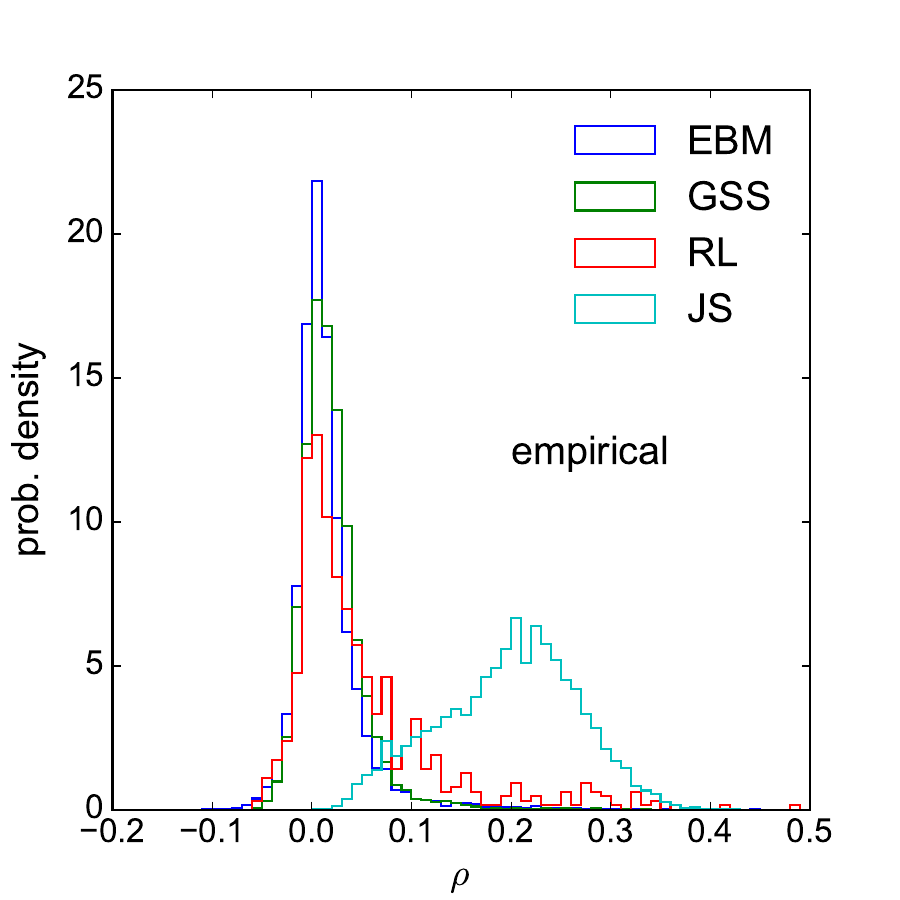}}
	\subfigure{\includegraphics[width=8cm]{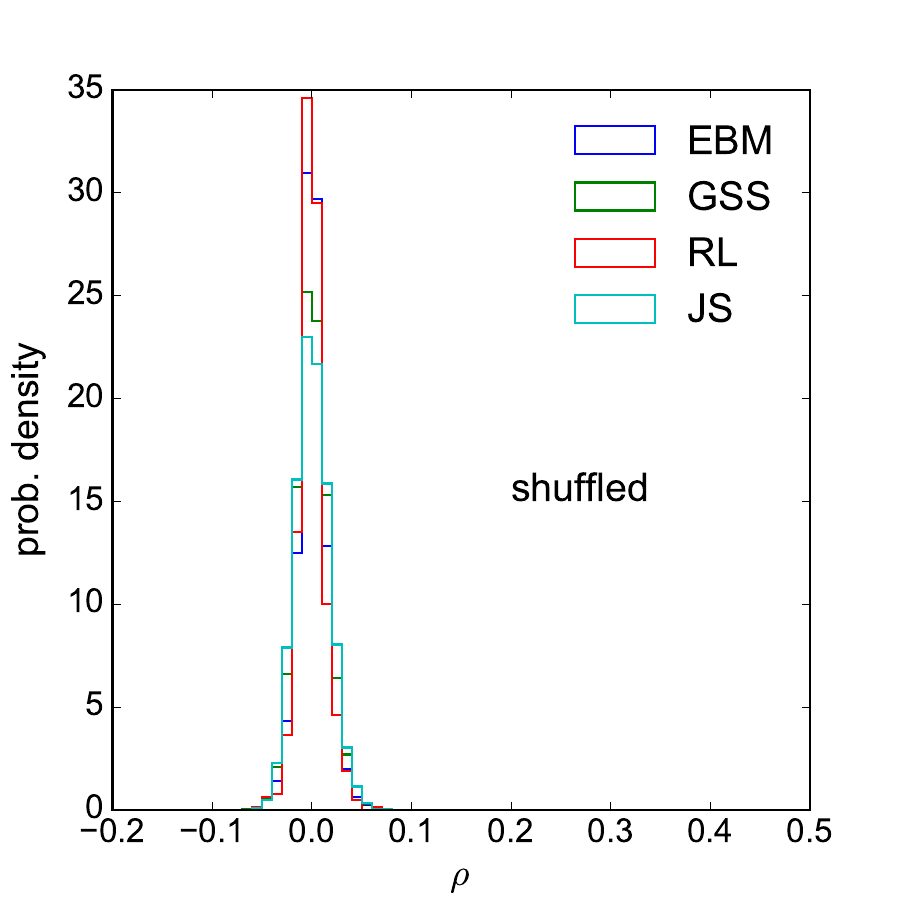}} 
	\caption{
	Distribution of feature-feature correlation $\rho$ for the empirical (left) and shuffled (right) versions of each of the four data sets (legend).
	Each histogram is normalized such that its integral is equal to 1, after being initially filled with $F(F-1)/2$ entries, where $F$ is the number of features in the respective data set,
	each entry corresponding to one pair $(k,l)$ of distinct features.
	For the normalization, the integral multiplies the bin content with the bin width $\delta\rho$ (the same for all histograms):
	the ordinate value of each bin is its relative frequency multiplied by a factor of  $1/\delta\rho$.
}
\label{CorHist}
\end{figure*} 

Another important aspect is the robustness of the LTCD-STCB curves of Fig.~\ref{FigEmpirDvC} when switching from one geographical region to another, which is illustrated here by Fig.~\ref{FigEmpirDvC_Geo}.
This is done by focusing on the two data sets which allow for division of the sample in terms of geographical regions, 
namely the Eurobarometer and the Religious Landscape.
Moreover, only the nominal-variable information in the Eurobarometer is being used, for reducing the computational time required to run the cultural dynamics model, as well as for illustrating the robustness of the results with respect to the sample of cultural variables that are used. 
The empirical and shuffled LTCD-STCB curves are being shown for 5 EU countries (left) and for 5 US states respectively (right).
Only one random curve is shown, because, for a specific data set, the country/state-level SCVs are defined with respect to the same cultural space,
which is fully determined by the types and ranges of variables in the empirical data, which are the same regardless of the sample of people.
Note that, for both data sets, the empirical and shuffled curves fall into clearly distinguishable bands.
The empirical curves are systematically above the shuffled ones, while being again close to the second diagonal.
This also suggests a geographical universality of the structural properties inherent in empirical data.

When confronted with these results, one thinks of unavoidable similarities between questions in the survey, which induce correlations between cultural features. 
Since these correlations are destroyed by the shuffling procedure, it is tempting to invoke them as an explanation for the discrepancy between an empirical LTCD-STCB curve and its shuffled counterpart. 
However, there is no reason to believe that such similarities are equally present in different empirical data sets, 
or that they are similarly distributed among the pairs of questions in the data set,
since different data sets rely on completely different sets of variables.
In fact, the measured feature-feature correlations $\rho^{k,l}$, defined via Eq.~\eqref{FFCor} are quite different across the four data sets used here.
This is illustrated by Fig.~\ref{CorHist}, which shows how the values of these correlations are distributed for the different empirical SCVs (left), 
while also showing, for comparison, the distributions for their shuffled counterparts (right), which, as expected, are strongly peaked around 0 
(the empirical and shuffled correlation matrices are shown in Figs.~\ref{CorMat} and~\ref{CorMat_Shuf} of Appendix Sec.~\ref{AppFFCors}).
The departure of the empirical distribution from its shuffled counterpart is clearly different across data sets, 
whereas the departure of the empirical LTCD-STCB curve from its shuffled counterpart is very similar across data sets, as shown in Fig.~\ref{FigEmpirDvC}.
Moreover, feature-feature correlations are typically small, given that any $\rho_{k,l}$ can take values within the $[-1,1]$ interval.
These are indications that the properties captured by the LTCD-STCB plot are not (or not exclusively) due to feature-feature correlations, 
and that additional information destroyed by shuffling (including higher-order correlations) plays an important role.
Such considerations enforce the idea that the observed properties are due to a more subtle, dynamical and universal mechanism. 

\section{Discussion}
\label{Disc}

The findings above stem from analyzing conventional social survey data in an unconventional way.  
Specifically, data from different sources is converted to empirical cultural states obeying a unified format,
which does not retain the meanings of the questions in the survey, nor the meanings of their associated answers, 
but just the frequency distribution of respondents in cultural space.
The LTCD and STCB quantities that are applied on the formatted data are also independent of the meanings of used variables and values, 
although highly sensitive to the distribution in cultural space.
This ``semantically-invariant'' nature of the analysis (invariance with respect to any relabeling of the cultural space that preserve all distances) 
is what allows one to potentially uncover universal properties in the structure of culture.

The results of the analysis suggest that there is something universal about how real people are distributed in cultural space. 
Empirical cultural states seem to induce a correspondence between LTCD and STCB that is highly robust across data sets, 
while significantly and consistently different from those induced by shuffled and random cultural states.
If empirical cultural states are regarded as partial snapshots of this dynamics,
the supposedly universal behavior could be seen as a consequence of general laws governing the dynamics of culture in the real world.
This rises the question of what these laws actually are: 
what is the mechanism giving rise to distributions in cultural space that are compatible with the above results? 
Answering this question might mean achieving a full understanding of cultural dynamics. 
If one thinks in terms of snapshots of culture, this is equivalent to finding a general theory of preference formation,
which is a fundamental challenge for the social sciences~\cite{Thompson}, 
with important implications for properly understanding decision making and economic behavior~\cite{Fehr_1, Fehr_2, Fehr_3}.
It appears that an important role for such a theory should be played by social influence,
as its role in the aggregation of individual opinions and the formation of collective opinions has been extensively studied~\cite{Galam_1, Galam_2, Galam_3}.
However, most of these studies focus on one-dimensional systems, while the empirical signatures presented are extracted from data with high dimensionality. 

From a theoretical perspective, bringing together multidimensional opinion spaces and the notion of social influence is achieved by Axelrod-like models of cultural dynamics.
Initializing the Axelrod dynamics with a random cultural state and studying the outcome goes along with
understanding the type of structure that social influence can dynamically give rise to, assuming a structureless initial state. 
If social influence alone is responsible for the structure observed in empirical data,
one would expect that an empirical cultural state is an intermediate outcome of the Axelrod dynamics.  
Thus, applying this dynamics to an empirical state would lead to an absorbing states that are statistically compatible with those obtained by applying the same dynamics to random states.
However, the analysis presented here, whose LTCD quantity incorporates full simulations of an Axelrod-like model, shows a clear and robust discrepancy between the random and the empirical states. 
This suggests that social influence is not enough for explaining the generic empirical structure highlighted by the analysis. 
Nonetheless, the Axelrod model used by the LTCD quantity is highly simplistic, disregarding geographical space, social networks, influence of media and other aspects that are present in the real world.
Moreover, the empirical cultural vectors correspond to individuals that are typically not interacting with each other directly in the real world, while they do so in the Axelrod model. 
Checking whether such considerations are sufficient for explaining the systematic discrepancies between random, shuffled and empirical cultural states is an interesting topic for further research. 
It these are not sufficient, more exotic model ingredients should be considered, such as cognitive processes~\cite{Sobkowicz_2} or logical constraints across cultural features~\cite{Friedkin}.

Contrary to the reasoning above, one can argue that the difference between the empirical and the shuffled regime of the LTCD-STCB analysis may simply be due to the presence of feature-feature correlations,
which in turn are supposedly due to ``design details'' of the social survey, having to do with certain questions being similar to each other.
Consequently, there would be no need to think about dynamical mechanisms responsible for the empirical structure.
However, the a-priori expectation is that design-induced correlations are relatively weak: 
collecting social survey data is expensive, so the survey should be designed such that it captures as much as possible of the relevant degrees of freedom, by minimizing the similarities among questions.
Moreover, remaining similarities should be specific to each data set, whereas the LTCD-STCB analysis gives highly similar results for different data sets.
To better illustrate this counterargument, feature-feature correlations were measured in Sec.~\ref{UnivStrProp} and explicitly shown to be specific to each social survey, 
which is compatible with the idea that they largely depend on ``design details'' -- see Appendix Sec.~\ref{AppFFCors} for more remarks along these lines.
In fact, feature-feature correlations can be seen as one of several manifestations of a non-uniform cultural space distribution, 
which is certainly also affected by a-priori, 
survey-dependent similarities between features, but arguably not in an essential way. 
It is also worth noting that one cannot say to what extent a correlation between two features is caused by an a-priori similarity between the two questions and to what extent it arises dynamically due to the combination of processes taking place in the real world. 
One can even argue that trying to disentangle the a-priori contribution is entirely meaningless, partly because the questions themselves are formulated by humans who interact with each other and with society.

Another aspect that this study pointed out is the strong dependence of social influence cultural dynamics and its final outcome on the initial cultural state.
This is dependence becomes manifest in the analysis presented in Sec.~\ref{UnivStrProp} as the systematic departure of the LTCD-STCB curve corresponding to empirical data from those corresponding to the shuffled and random counterparts.
confirming and expanding the results of Refs.~\cite{Valori, Stivala}.
The dependence on initial states is rarely studied in the literature on cultural/opinion dynamics.
A notable exception is Ref.~\cite{Galam_4}: 
upon analysing the Metropolis dynamics of the Ising model using an analytic technique developed in the context of opinion dynamics, 
a regime is found that allows for several, qualitatively different equilibrium states to be reached, depending on the initial configuration. 
It is also worth noting that, for studying the Axelrod model, Ref.~\cite{Castellano_2} is using a non-uniform distribution in cultural space for randomly generating its initial states.  
Still, it is a distribution that can be factorized as a product of Poisson, feature-level distributions, encoding no structure in addition to that entailed by the feature-level non-uniformities. 
Refs.~\cite{Valori, Stivala} also suggest that initial state dependence can be understood in terms of an ultrametric appearance of real cultural data, 
observation which Ref.~\cite{Stivala} exploits for developing static models of cultural states characterised by a hierarchical organization in cultural space. 
Although this line of reasoning has not been used here, it should be further explored by future work. 

Defining a (probabilistic) model of cultural states would be equivalent to specifiying a cultural space distribution, the model being more realistic when the empirical data is better representative of this distribution. 
Such future research is further motivated by the robust behavior identified by this study, and by the observation that the three types of cultural states appear to roughly fall into three equivalence classes, 
in terms of the shapes of the associated LTCD-STCB curves. 
The purpose would be to design a model that generates artificial SCVs falling under the empirical equivalence class.
Once the model is in place an properly tunned, the anlysis of SCVs can in principle be extended to regimes that are not empirically accessible, due to limitations on $F$ and $N$. 
This should allow for more detailed, statistial physics work to be done in relation to the phase transitions described in Sec.~\ref{LTCDaSTCB} and Sec.~\ref{UnivStrProp},
such as finite-size scaling analysis and measurement of critical exponents. 
One might also achieve a better understanding of the extent to which the notion of self-organized criticality is important, 
by analysing the distribution of cluster sizes in cultural space for interesting $\omega$ values.
At this point, this is highly speculative, based on the apparent complementarity between the LTCD and STCB transitions for empirical data, 
as well as on accepting that real-world systems are favourable for both long-term cultural diversity and short-term collective behavior.   
One can object by arguing that the shape of the LTCD and STCB transitions are sensitive to the exact mix of ingredients going in evaluating the two quantities 
-- for instance, one can imagine using a more sophisticated Axelrod-type mode for evaluating LTCD. 
However, in the manner used here, LTCD and STCB are defined in a very similar, minimalistic way:
adding more ingredients, such as geographical space and social networks, should be done in parallel for both quantities. 
It is plausible that additional ingredients would alter the two transitions in the same way, such that the relationship between LTCD and STCB is preserved.

\section{Conclusion}
\label{Conc}

This study is an additional step towards understanding the dependence of social-influence cultural dynamics on the initial cultural state state.
At the same time, it provides insights about the structure inherent in empirical cultural data by means of its effect on cultural dynamics,
evaluated by the LTCD quantity, 
conditional on its effect on shorter time-scale opinion dynamics, 
evaluated by the STCB quantity.
It turns out that the LTCD-STCB combination, together with comparisons between empirical data and randomized counterparts, 
suggest the existence of universal properties characterising how real people are distributed in cultural space.
These properties seem to be present in spite of the variabilities of the feature-feature correlation matrix across data sets.
Further work is needed to understand in more depth the nature and implications of these properties. 

{\bf Acknowledgements:} 

The authors are grateful to Maroussia Favre for her thoughtful comments on previous versions of this manuscript.
AIB also acknowledges discussions with Andreas Flache, Gerard 't Hooft, Michael M\"{a}s, Michael Thompson, Marco Verweij and Jorinde v.d. Vis.
DG acknowledges financial support from the Dutch Econophysics Foundation (Stichting Econophysics, Leiden, the  Netherlands).
This work was also supported by the Netherlands Organization for Scientific Research (NWO/OCW).

{\bf Author contributions:} 

AIB and DG designed the research. AIB wrote the computer code. LT carried out the preliminary data formatting and analysis. AIB carried out the final data formatting and analysis. AIB and DG wrote the manuscript. 

\appendix

\section{Empirical data formatting}
\label{AppEmpDat}

This section explains various details concerning the formatting of empirical data. 
As previously mentioned, four data sets were employed, each of which was collected by different entities, for different purposes and in different formats. 
In order for the analysis and modeling conducted here to be carried out consistently, the important information had to be extracted from each data set and expressed in one, unified format. 
Essentially, this format dictates that each data set has to provide a certain number of ordinal features and a certain number of nominal features, 
where each feature has a certain number of possible traits (the range $q$ of the feature),
and that the traits of every individual in the data set are recorded with respect to all these features. 
This unified format can be effectively thought of as a table of traits, where the rows correspond to the features and the columns correspond to the individuals. 
There are various challenges involved when converting the data into this format. 
It is worth explaining first the challenges that are more generic, relevant for several data sets
and scond the challenges specific to each data set. 

One of the difficulties consists in deciding, for each variable, whether it should be used as cultural feature or not.
The following is a (not entirely exhaustive) list of types of variables which are worth mentioning in this regard:
\begin{itemize}
	\item demographic variables, such as those encoding ``age'', ``place of residence'' or ``ethnicity'' are discarded, as they do not record subjective human traits;
	\item	certain variables, that were not seen as demographic variables by the survey authors, are also discarded if they recorded information about something that is too much in the respondent's past, 
				or about something that cannot be easily related to subjective preferences, opinion, values, beliefs or behavioral tendencies that can be conceivably altered via social influence in a reasonably easy way;
				often, the boundary between what is subjective and what is objective not clear;
				nonetheless, one can strive to make these decision consistently at the level of every data set, which is what was done here;
	\item there are questions that ask opinions with respect to something that is differently defined for different people in the survey, such as: 
				``how satisfied you are about how the the economy of this country is going recently?'' -- if there are people from different countries in the data set, 
				or ``how satisfied you are with your life?''; these questions are also discarded;
	\item questions asking the respondent to self-evaluate a certain, personal trait, such as ``would you say about yourself that you are more conservatory or liberal on political affairs'', are retained,
				assuming that the respondent mostly self-evaluates, in a reasonably objective way, a personal (subjective) trait, rather than expressing a subjective opinion about the personal trait;
	\item certain variables containing relevant information are also discarded if, due to the survey format, they can only be answered when certain answers are given to other variables,
				or if the set of possible answers explicitly depends on answers given to other questions,
				regardless of whether these "other" variables themselves are selected or not;
				including such variables would introduce inconsistencies in the encoding of cultural vectors, the definition of cultural distance and the shuffling and randomization procedures. 
\end{itemize}

The variables that are retained for further analysis need to be encoded either as nominal or ordinal cultural features.
Deciding between the two encoding options was done here using the following criterion: 
if there are more than two possible answers that are not ``neutral'' (see next paragraph) and they can all be conceivably ordered along the real axis, then the variable is encoded as ordinal;
if, instead, there are only two answers (typically ``Yes'' and ``No'') in addition to the neutral ones, or if the non-neutral answers cannot be ordered along the real axis in a consistent way, then the variable is encoded as nominal.

Most variables retained from the data sets also allow for one or more ``neutral'' answers (often called ``missing values'' in social science research, although this term usually is somewhat more general).
These are usually labeled as ``Don't know'', ``Refused'' or ``Not Answered''. 
For further analysis, these neutral answers are merged (if more than one are present).
If the variable is to be encoded as nominal, neutral answers are mapped to one, additional cultural trait, side-by-side with traits originating from non-neutral answers. 
If the variable is to be encoded as ordinal, they are mapped to the middle of the ordinal scale 
-- if there is an even number of possible answers, for each person, the choice is randomly made between the two answers closest to the middle of the scale.

Note that some data sets (GSS and EBM below) formally allow for another type of answer, labeled as ``IAP'' or ``INAP'' (inapplicable), 
which is here regarded as separate from neutral answers (although in social science research they are often all placed under the ``missing values'' umbrella term).
IAP values are recorded, for certain respondents, when answers to a specific question are not expected from those respondents, 
for reasons having to do with the design of the survey.
This happens for question that are only asked conditionally on answers given before.
However, as mentioned above, these conditional variables are anyway discarded. 
Similarly, IAP values are also recorded for questions that are only asked to a certain sub-sample of the people, 
although not being conditional on some other question, in which case those questions are either removed or, if the sub-sample is large enough, 
the formatting is restricted to it. 
Finally, IAP values are also recorded for split-ballot or split-form variables (see GSS and EBM explanations below), in which case specific procedures are followed, 
which effectively discard all IAP answers before further analysis.
Thus, regardless of how exactly they occur, one does not need to map IAP answers to any trait, as they are all filtered out as a consequence of other formatting rules.
Note that for the RL data set, although IAP answers are not explicitly mentioned anywhere, this could have been the case, 
since there are questions that are conditionally asked on other questions -- instead of IAP answers, system-missing values are present in the SPSS file, typically marked by the ``.'' dot character. 

First, this study made use of the {\bf Jester 2 (JS)} data set~\cite{JS},
which consists of online ratings of jokes collected between November 2006 and May 2009.
There are around 1.7 million continuous ratings (on a scale from -10.00 to +10.00) of 150 jokes from 59,132 users.
For most users however, of the 150 jokes, only 128 are provided as items to be rated, as the other 22 were eliminated at a certain point in time. 
For this study, each of the 128 items is converted into an ordinal feature with 7 traits (by splitting the $[-10,10]$ interval into 7 bins of equal size, 
while assuming that everything falling within one bin constitutes the same answer).
Moreover, only the 2916 users that had rated all items were retained for further analysis -- 
although this introduces some bias in the sample, one can argue that it is desirable to focus on individuals that have rated everything, 
as this is an indication of commitment on the respondent's side.
	
Second, the research used the {\bf Religious Landscape (RL)} data set~\cite{RL}, 
which consists of opinions and attitudes on various religious topics, but also on various political an social issues. 
These data were collected in 2007 via telephone interviews from all states of USA -- this study only used the data obtained from the continental part of the USA (without Hawaii and Alaska). 
There are multiple questions asking about the religious affiliation of respondents, which were all discarded.
This is partly based on the assumption that religious affiliation is closer to a demographic variable than to a feature that can be easily altered via social influence,
partly based on the very large number of answers and the nested, hierarchical nature of how they are organized.
For this study, 36 cultural features were constructed (18  nominal and 18 are ordinal), for a number of 35558 respondents.
  
Third, the research used the {\bf Eurobarometer 38.1 (EBM)} data set~\cite{EBM}, 
which consists of opinions on science, technology, environment and various EU political issues (mainly related to the open market and the economy). 
The data were collected during November 1992, from 12 countries of the EU, via face-to-face interviews.
In this survey, there are several blocks of ``coupled'' variables  which are all discarded: 
within each block, there are explicit internal constraints on how answers can be given (such as answering ``yes'' to at most 3 questions out of 8 that are available), 
which do not allow for a consistent encoding as a set of nominal or ordinal features. 

Another challenge when formatting the EBM data set is posed by the split-ballot procedure: 
the sample of people is split into 2 ballots, and certain questions are asked in slightly different versions (small differences in formulation, answers listed in different orders etc.) to the two ballots, 
while both versions are present in the SPSS file for all individuals -- for every respondent, an IAP answer is recorded for the version that is not used for that respondent.
The most meaningful approach is to merge the two versions and eliminate all IAP answers -- if both versions are kept, strong structural artifacts arise in the matrix of cultural distances~\cite{Stivala}.
Most of the split ballot variables are encoded as ordinal and have the same range (same number of non-neutral answers) in both versions, such that a one-to-one correspondence can be made, similarly to Ref.~\cite{Stivala}.			
Some of them are still ordinal but have different ranges in the two versions. 
In all these cases, there is a difference of only one trait among the two versions, such that one range is an even number while the other is odd. 
In this case, the odd version is kept for the merging, which guarantees the existence of a middle trait to which all neutral answers can be directly assigned.	
The non-neutral answers from the even version are mapped to the closest answers in the odd version, 
in terms of the distance from the lowest-value answer, assuming that the distance between the lowest-value and highest-value answers is the same in the two versions (consistent with the definition of cultural distance in Eq.~\eqref{CultDist}).
There is one split ballot variable which is encoded as nominal, 
in which case the difference consists in a second question being asked for one of the ballots, which is simply discarded.
After all the formatting, 144 cultural features are constructed from this data set (54 nominal and 90 ordinal),
for a number of 13026 respondents. 

Fourth, the study used the {\bf General Social Survey (GSS)} data~\cite{GSS}, collected during 1993 in the USA via face-to-face interviews.
The overall scheme of how questions are asked to respondents is arguably more complicated than for the EBM data set.
First, there is a split-form procedure involved, which is equivalent to what is called ``split-ballot'' in the case of EBM:
the respondents are split into two groups, with certain questions being asked in two, slightly different versions.
All these questions are ordinal and have the same ranges in the two forms; they are handled like in the case of EBM. 
Independently of the split-form procedure, there is another procedure called ``split-ballot'', which is methodologically somewhat different:
the sample of respondents is split in 3 ballots (A,B,C), while some questions are only asked to 2 of the 3 ballots (A and B, B and C or A and C).
This is handled by discarding the questions asked to only 2 of the 3 ballots.
Independently of the split-ballot and split-form procedures, there is a set of questions, also used within the International Social Survey Program (ISSP),
which are not asked to a small fraction of respondents (49 out of 1608 respondents).
This is handled by discarding the 49 people not exposed to the ISSP questions. 
All in all, 133 cultural features are constructed from the GSS data (8 nominal and 125 ordinal), for a number of 1559 respondents. 
		
\section{Feature-feature correlations}
\label{AppFFCors}
		
\begin{figure*}
\centering
	\subfigure{\includegraphics[width=8cm]{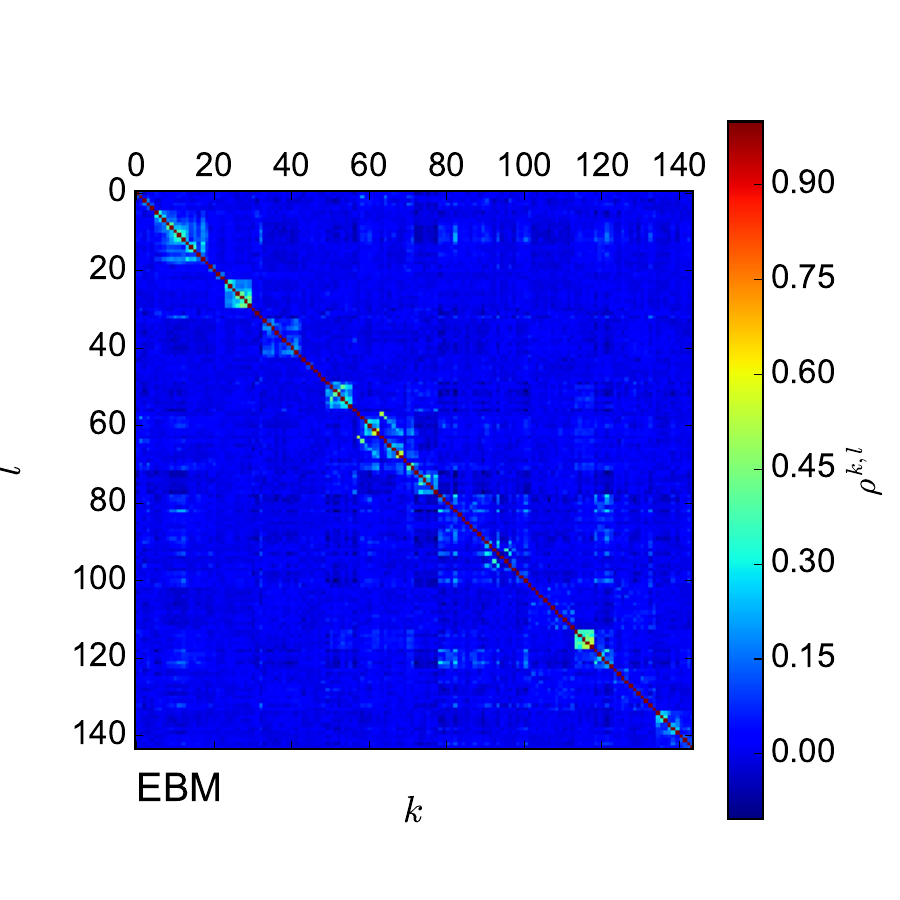}}
	\subfigure{\includegraphics[width=8cm]{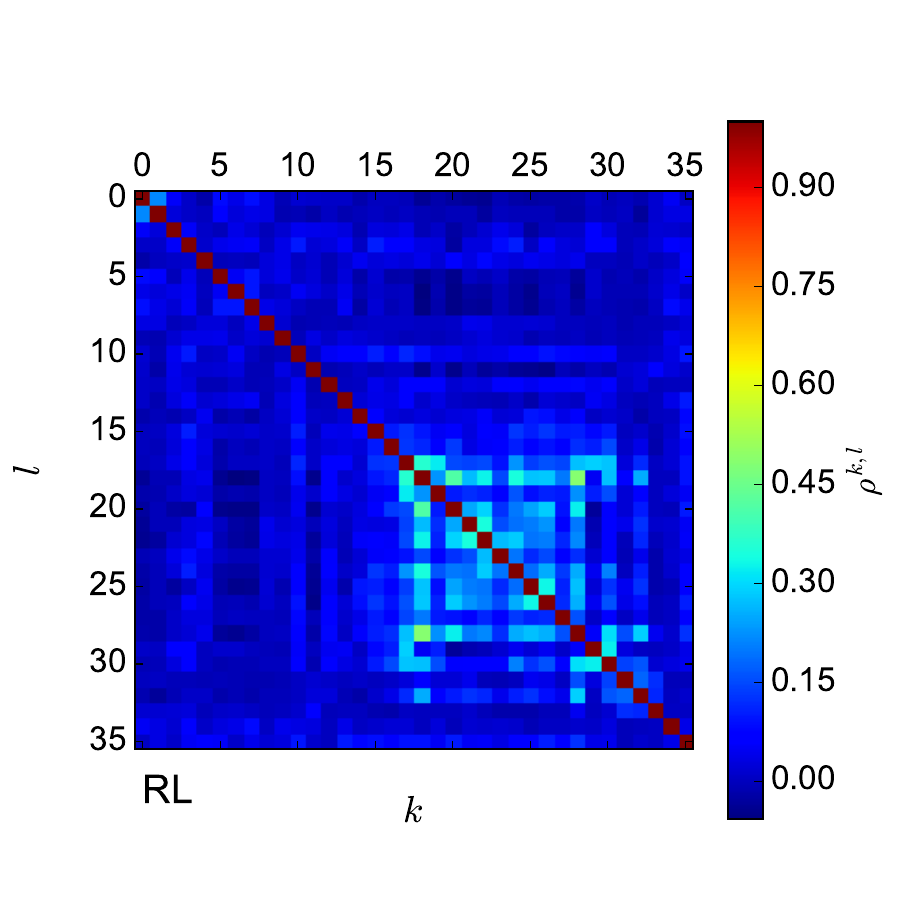}} \\ \vspace{-0.3cm}
	\subfigure{\includegraphics[width=8cm]{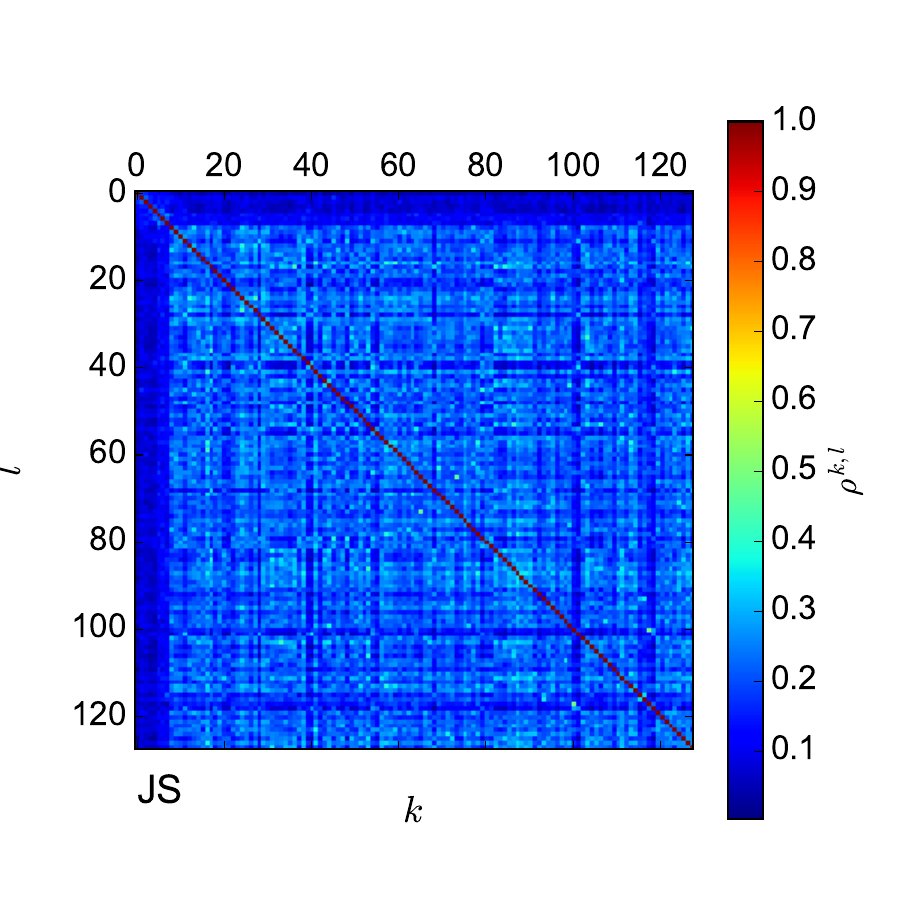}}
	\subfigure{\includegraphics[width=8cm]{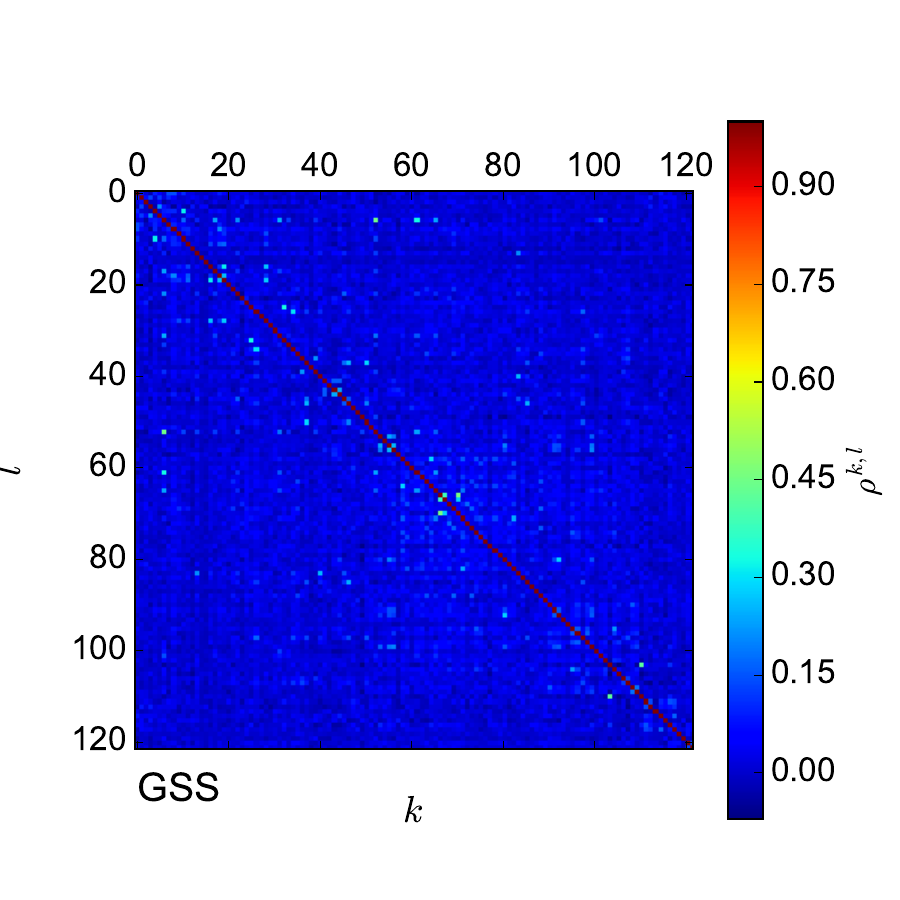}} 
	\caption{
	Matrix of feature-feature correlations in empirical sets of $N=500$ cultural vectors obtained from the four sources: Eurobarometer (EBM), Religious Landscape (RL), Jester (JS) and General Social Survey (GSS). 
	Each grid point shows the correlation $\rho^{k,l}$ between cultural features $k$ and $l$.
}
\label{CorMat}
\end{figure*} 		

\begin{figure*}
\centering
	\subfigure{\includegraphics[width=8cm]{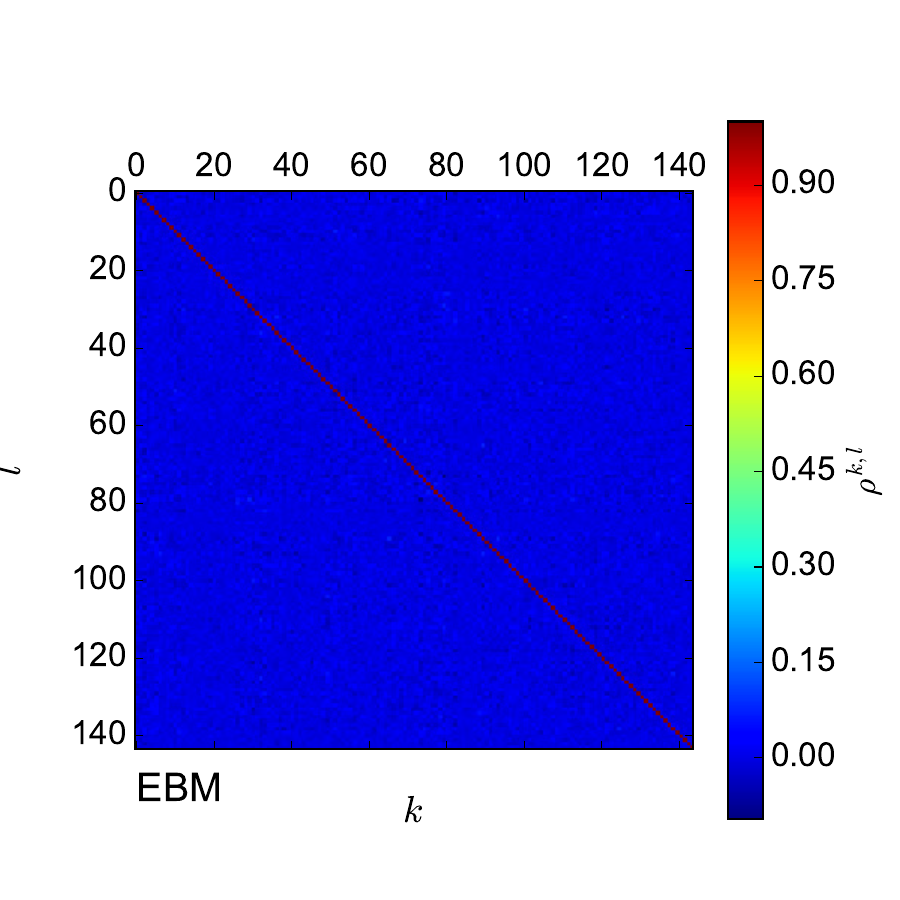}}
	\subfigure{\includegraphics[width=8cm]{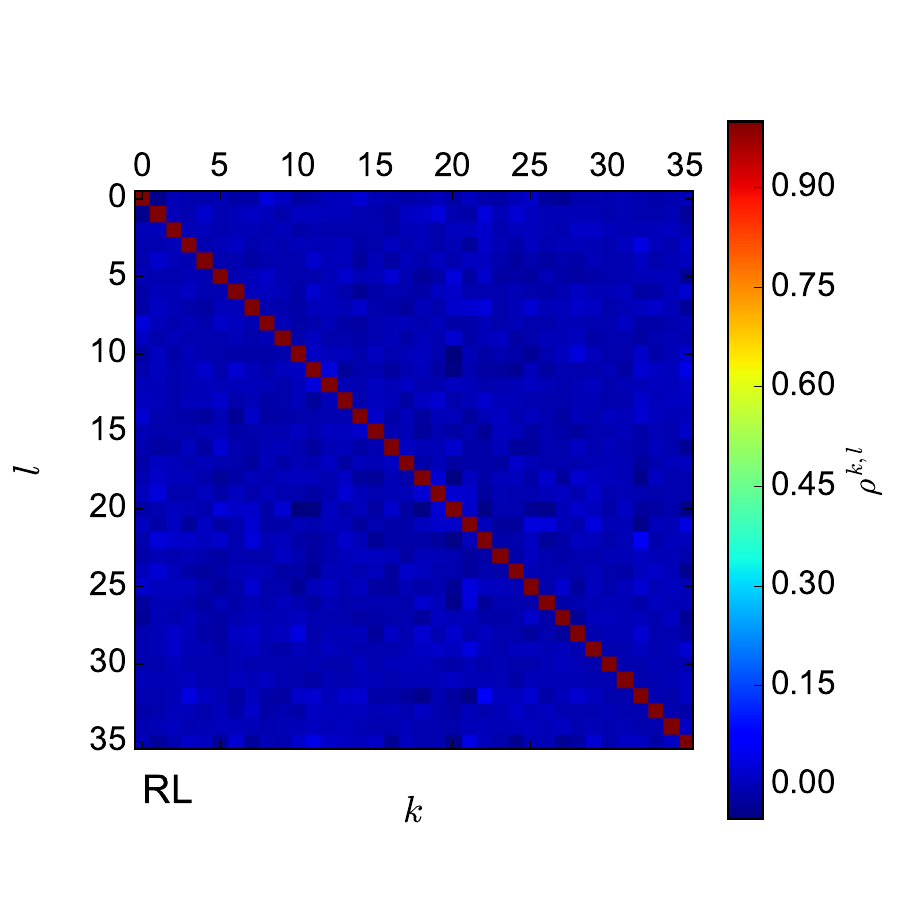}} \\ \vspace{-0.3cm}
	\subfigure{\includegraphics[width=8cm]{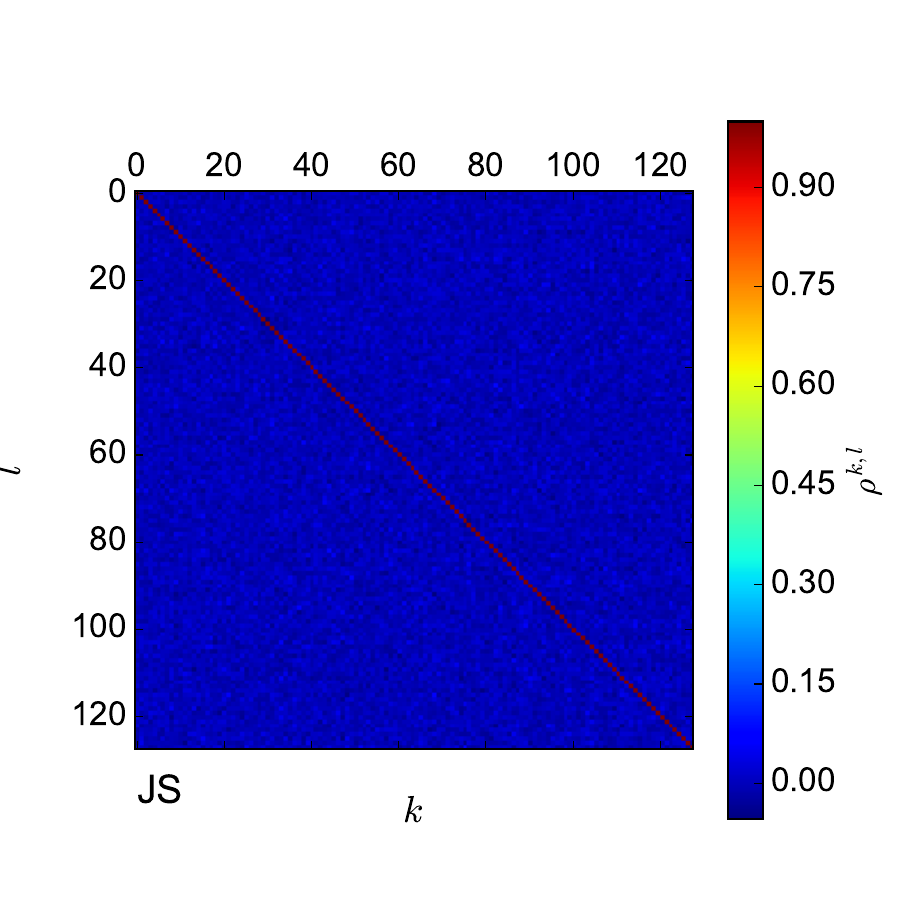}}
	\subfigure{\includegraphics[width=8cm]{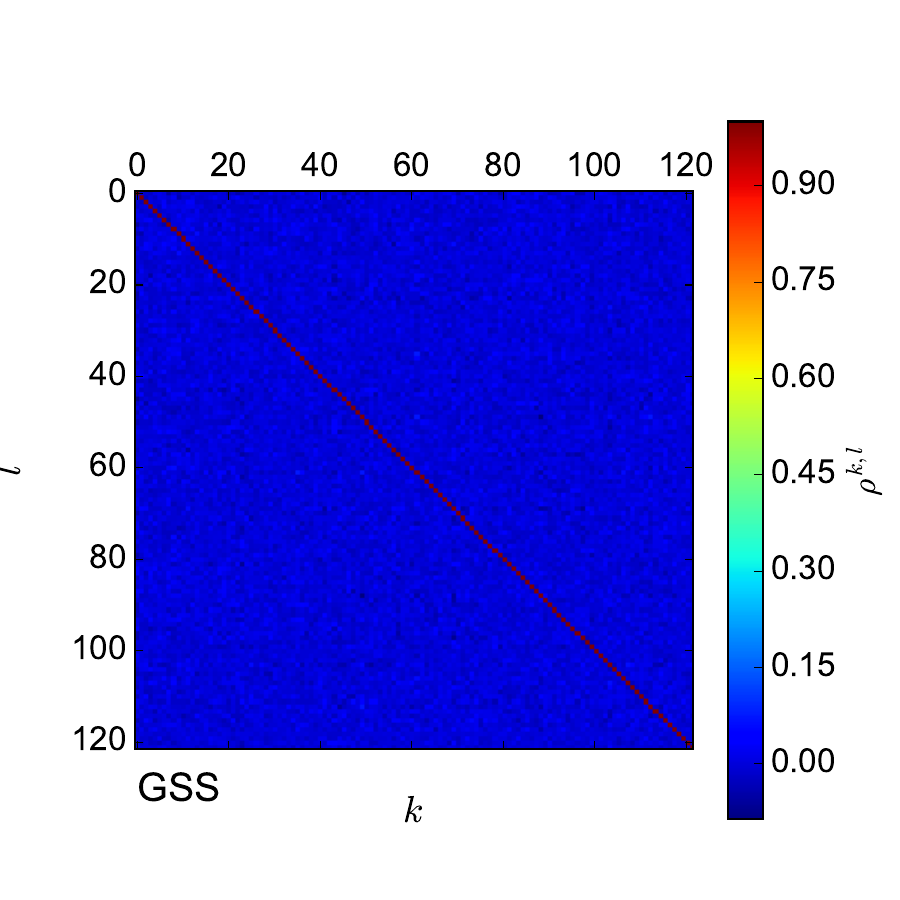}} 
	\caption{
	Matrix of feature-feature correlations in shuffled sets of $N=500$ cultural vectors corresponding to the four empirical sources: Eurobarometer (EBM), Religious Landscape (RL), Jester (JS) and General Social Survey (GSS). 
	Each grid point shows the correlation $\rho^{k,l}$ between cultural features $k$ and $l$.
}
\label{CorMat_Shuf}
\end{figure*} 		

This section illustrates in detail the correlations between cultural features, computed according to Eq.~\eqref{FFCor}.
The feature-feature correlation matrices of the four empirical SCVs are shown in Fig.~\ref{CorMat}, while those of the four shuffled counterparts are shown in Fig.~\ref{CorMat_Shuf}.
The ordering or rows and columns is consistent with the actual ordering of questions in the four data sets.
This leads to a partial block-diagonal aspect of the matrices associated to the Eurobarometer and Religious Landscape data sets, for which questions that deal with similar topics tend to appear next to each other.
Note that, empirical correlations rarely show strong deviations from their shuffled counterparts.
Interestingly, the largest level of correlation is visible for the Jester (JS) data set, which is certainly the least expensive to collect, since respondents provide their answers online, via an automated platform. 
Moreover, the second-largest level of correlation is present in the Religious Landscape (RL) data set, which is arguably the second-least expensive to collect, since it relies on telephone interviews, 
while the other two data sets rely on face-to-face interviews.
This is supports the idea that such correlations are survey specific, that they tend to be minimized by survey design 
and that they are not responsible for the generic structural properties identified by this study.
There is a clear discrepancy between the Eurobarometer correlation matrix shown here and that shown in the Supplementary Information of Ref.~\cite{Valori}. 
However, the current study used a different, much more rigorous procedure of formatting the empirical data.

\bibliographystyle{unsrt}
\bibliography{Paper}{}

%\begin{thebibliography}{2}

%\end{thebibliography}

\end{document}